\documentclass{svproc}

\usepackage{graphicx}
\usepackage{times}
\usepackage[body={16cm, 24cm}]{geometry}

\begin{document}
\mainmatter   
\title{ Inclusive breakup reaction of a two-cluster projectile on a two-fragment target: A genuine four-body problem}

\titlerunning{Incomplete breakup}  % abbreviated title (for running head)

\author{M. S. Hussein\inst{1,2,3} \and C. A. Bertulani\inst{4} \and B. V. Carlson\inst{3} \and T. Frederico\inst{3}}
\institute{Instituto de Estudos Avan\c{c}ados, Universidade de S\~ao Paulo, Caixa Postal 72012, 05508-970 S\~ao Paulo, SP, Brazil
\and Instituto de F\'{\i}sica, Universidade de S\~ao Paulo, Caixa Postal 66318, 05314-970 S\~ao Paulo, SP, Brazil
\and Departamento de F\'{\i}sica, Instituto Tecnol\'{o}gico de Aerona\'{u}tica, DCTA, 12.228-900 S\~ao Jos\'{e} dos Campos, SP, Brazil
\and Department of Physics and Astronomy, Texas A\&M University-Commerce, Commerce, TX 75429-3011, USA}
\maketitle

 \noindent

\begin{abstract} 
We develop a four-body model for the inclusive breakup of two-fragment halo projectiles colliding with two-fragment targets. In the case of a short lived projectiles, such as halo nuclei, on a deuteron target, the model allows the extraction of the neutron capture cross section of such projectiles. We supply examples.
\end{abstract}

\section{Introduction}
The ongoing research on the reaction of radioactive nuclei has supplied us with invaluable information about the structure of nuclei near the drip line. Further, they produced important information on capture reactions and other direct reactions needed to fill the gaps in the chain of reactions in the r and s processes in astrophysics. The neutron capture reactions referred to above involve capture by stable nuclei. Neutron capture reactions on radioactive nuclei, especially near the drip nuclei are not available. A possible way to obtain these cross sections is through indirect hybrid reactions. One such method is the Surrogate Method \cite{Escher2012}. So far this method was mostly used to obtain neutron capture cross section of fast neutrons by actinide nuclei for use in research in fast breeder reactors. Recently, the Surrogate Method was proposed to obtain the neutron capture cross section of radioactive nuclei \cite{Escher2018}. A recent review gives an account of the (d, p) inclusive breakup reaction which is the basis of the Surrogate Method \cite{Potel2017}. The theory employed  for this is the Inclusive Nonelastic Breakup  (INEB) Reaction theory \cite{IAV1985,UT1981,Ichimura1982,HM1985,Austern1987}. In this contribution  we report on recent work that extends the application of the INEB to the case of capture by a radioactive target or projectile. In our approach we consider first the three-body case of a non-cluster  projectile interacting with a two-cluster target, such as the deuteron. In this case the reaction is a neutron pickup. Through the measurement of the inclusive proton spectrum one is able to extract the neutron capture cross section. This cross section is not the free capture cross section as several factors come into play owing to the fact that the neutron is bound in the deuteron.  The second case we consider is the four-body one involving three-cluster projectile and no-cluster target \cite{CFH2017}. In this contribution we propose an extension of the theory of \cite{CFH2017} to the case of a two-fragment projectile on a two-fragment target. One such reaction involves the one proton halo nucleus $^8$B,  $^8$B + d $\rightarrow$ p + $^9$B or p + ($^7$Be + d). So the inclusive proton spectrum will exhibit two groups a low proton energy one associated with the incomplete fusion $^7$Be + d and a higher proton energy group connected with the capture reaction. We also consider the one-neutron halo projectiles on the deuteron target, $^{11}$Be + d and the $^{19}$C + d. Our work reported here should be useful to assess the applicability of the INEB theory to isotopes such as $^{135}$Xe whose lifetime is 9.8 hours, which is a notorious nuclear reactor poison as its thermal neutron capture cross section is huge, 2.5$\times 10^6$ barns. Several other nuclei exhibit very large thermal neutron capture cross sections \cite{Mug2003}, whose explanation was attempted in \cite{CHK2016}. Our aim is to use $^{135}$Xe as a benchmark to test the inclusive proton spectrum in a reaction of the type d + $^{135}$Xe $\rightarrow$ p + $^{136}$Xe. 
\section{Two-fragment projectile on a one-fragment target nucleus}
We will consider the scattering of a radioactive projectile with a two-cluster target (deuteron). Let us take $^{135}$Xe as an example. Its life time is 9.6 hours (very long). The reaction we want to describe is $^{135}$Xe + d $\rightarrow$ p + $^{136}$Xe. A pickup reaction. The spectrum of the protons is measured, and the theory for this inclusive reaction is available. The quantity which is extracted from the measurement and the analysis is the total reaction cross section n + $^{135}$Xe $\rightarrow$ $^{136}$Xe, the neutron capture reaction. The cross section is given by the Austern or Hussein-McVoy (HM) expression
\begin{equation}
\frac{d^{2}\sigma_{p}}{dE_{p}d\Omega_p} = \rho_{p}(E_p) \hat{\sigma}_{R}^{n},
\end{equation}
where $\hat{\sigma}_{R}^{n}$ is the medium-modified total reaction cross section of the process n + A,
\begin{equation}
\hat{\sigma}_{R}^{n} = \hat{\sigma}_{R}(n + A).
\end{equation}

More explicitly, the Inclusive Nonelastic Breakup theory gives for the reaction a + A $\rightarrow$ b + (x + A)
\begin{equation}
\frac{d^{2}\sigma^{\rm INEB}_b}{dE_{b}d\Omega_{b}} = \hat{\sigma}_{R}^{x}  \ \rho_{b}(E_b), \label{inebI}
\end{equation}
where $\hat{\sigma}_{R}^{x}$ is the total reaction cross section of the interacting fragment, $x$, and 
\begin{equation}
\rho_{b}(E_b)  \equiv {d\textbf{k}_{b}\over (2\pi)^3}{1\over dE_{b}d\Omega_{b}}  = {\mu_{b}k_{b}\over (2\pi)^{3}\hbar^2}
\label{density}
\end{equation} 
is the density of state of the observed, spectator fragment, $b$. The reaction cross section $\hat{\sigma}_{R}^{x}$ is given by \cite{Hussein1984}
\begin{equation}
\hat{\sigma}_{R}^{x} = - \frac{k_x}{E_x}\langle\hat{\rho}_{x}(\textbf{r}_x)\left|W_{x}(\textbf{r}_x)\right|\hat{\rho}_{x}(\textbf{r}_x)\rangle  ,
\label{sigR}
\end{equation}
where $W_{x}$ is the imaginary part of the complex optical potential, $U_{x}$, 
of the interacting fragment, $x$, in the field of the target, A. The source function $\hat{\rho}_{x}(\textbf{r}_x)$ is the overlap of the distorted wave of the interacting fragment, $x$, and the total wave function of the incident channel. In the DWBA limit of the this latter wave function and using the post form of the interaction, $V_{xb}$, the source function  in the HM approach \cite{HM1985}, is just 
\begin{equation}
\hat{\rho}_x(\textbf{r}_x) = (\chi^{(-)}_{b}|\chi^{(+)}_{a}\Phi_{a}>(\textbf{r}_x) .\label{rho1}
\end{equation}
In the IAV theory \cite{IAV1985}, based on the post form of the interaction, $V_{xb}$, the source function contains a Green's function referring to the propagation of $x$, 
\begin{equation}
\hat{\rho}_x(\textbf{r}_x) = \frac{1}{E_x - U_x + i \varepsilon} (\chi^{(-)}_b|V_{xb}|\chi^{(+)}_{a}\Phi_{a}\rangle \label{rho2} .
\end{equation}

The cross section in Eq.(\ref{sigR}), according to the Hussein-McVoy model \cite{HM1985}, can be decomposed into partial waves giving
\begin{equation}
\frac{E_{x}}{k_x}\hat{\sigma}_{R}^{x} = \int d\textbf{r}_{x} |\hat{S}_{b}(\textbf{r}_x)|^{2} W(\textbf{r}_x)|\chi^{(+)}_{x}(\textbf{r}_{x})|^2 ,
\label{sigRpar}
\end{equation}
where  
\begin{equation}
\hat{S}_{b}(\textbf{r}_x) \equiv \int d\textbf{r}_{b}  \langle\chi^{(-)}_{b}|\chi^{(+)}_{b}\rangle(\textbf{r}_b)\Phi_{a}(\textbf{r}_b, \textbf{r}_x), 
\label{S}
\end{equation}
and $\Phi_{a}(\textbf{r}_b, \textbf{r}_x)$ is the internal wave function of the projectile which carries the observed spectator fragment, b.
The above formalism has recently been employed to calculate the (d, p) inclusive proton spectrum in (d, p) reactions 
\cite{Potel2015,Ducasse2015,Moro1-2015,Moro2-2015,Carlson2015,Moro2017}.

In applying the above formalism to the reaction involving the deuteron as a projectile and $^{135}$Xe as the target, or vice versa, one is reminded once again of the lifetime of the latter, 9.8 hours. So there is the practical question which of these two reactions is feasible. In any case the final result in either case is the medium-modified total reaction cross section of the system n + $^{135}$Xe. The capture cross section is the difference between this cross section and the contributions of other direct reactions, such as inelastic excitation of $^{135}$Xe. In passing we remind the reader once again that in free space the thermal neutron capture cross sections of several nuclei is abnormally large~\cite{Mug2003,CHK2016}.

\section{Inclusive Non-Elastic Breakup Reactions of three fragment projectiles }

Recently we have developed the theory of INEB involving a three-fragment projectiles, $a = b + x_1 +x_2$, such as $^9$Be = $^4$He + $^4$He + n and Borromean nuclei such as $^{11}$Li = $^9$Li + n + n, The cross section for this four-body process, $ b + x_1 + x_2 + A$, where $b$ is the observed spectator fragment and $x_1$ and $x_2$ are the interacting participants fragments, is
\begin{equation}
\frac{d^{2}\sigma^{INEB}_b}{dE_{b}d\Omega_{b}} = \rho_{b}(E_b)\sigma_{R}^{4B},
\end{equation}
\begin{equation}
\sigma_{R}^{4B} = \frac{k_a}{E_a}\left[\frac{{E_{x_1}}}{{k_{x_1}}}\sigma_{R}^{x_1} + \frac{{E_{x_2}}}{{k_{x_2}}}\sigma_{R}^{x_2} + \frac{E_{CM}({{x}_1},{{ x}_2})}{(k_{{x}_1}+ k_{{x}_2})} \sigma_{R}^{3B}\right],\label{CFH-R}
\end{equation}
where, the form of the reaction or fusion cross section as derived in \cite{Hussein1984} is used,
\begin{equation}
\sigma_{R}^{x_1} = \frac{{k_{x_1}}}{E_{{x_1}}} \langle \hat{\rho}_{{x}_1, {x}_2}|W_{x_1}|\hat{\rho}_{{x}_1, {x}_2}\rangle, \label{sigma_1}
\end{equation}
\begin{equation}
\sigma_{R}^{x_2} = \frac{{k_{x_2}}}{E_{{x_2}}} \langle \hat{\rho}_{{x}_1, {x}_2}|W_{x_2}|\hat{\rho}_{{x}_1, {x}_2}\rangle, \label{sigma_2}
\end{equation}
and,
\begin{equation}
\sigma_{R}^{3B} = \frac{(k_{{x}_1}+ k_{{x}_2})}{E_{CM}({{x}_1},{{ x}_2})}\ \langle \hat{\rho}_{{x}_1, {x}_2}|W_{3B}|\hat{\rho}_{{x}_1, {x}_2}\rangle ,\label{sigma_3B}
\end{equation}
is a three -body, $x_1 + x_2 + A$, reaction cross section. The energies of the different fragments are defined through the beam energy, since the projectiles we are considering are weakly bound and thus the binding energy is marginally important in deciding the energies of the three fragments. Thus, e.g., $E_{{x_1},Lab} = E_{a, Lab}(M_{{x_{1}}}/M_a)$, where by $M_{a}$ and $M_{{x_1}}$ we mean the mass numbers of the projectile and fragment $x_1$, respectively. The three-body source function, $\hat{\rho}_{{x}_1, {x}_2}$, is a generalisation of the two-body source function in Eqs. (\ref{rho1},\ref{rho2}),
\begin{equation}
\hat{\rho}_{x_1, x_2} (\textbf{r}_{x_{1}}, \textbf{r}_{x_{2}})= 
(\chi^{(-)}_{b} (\textbf{r}_b)|\chi^{(+)}_{a}(\textbf{r}_{b}, \textbf{r}_{x_{1}}, \textbf{r}_{x_{2}})\Phi_{a}(\textbf{r}_{b}, \textbf{r}_{x_{1}}, \textbf{r}_{x_{2}})\rangle .
\end{equation}

The cross sections $\sigma_{R}^{x_1}$ and $\sigma_{R}^{x_2}$ are the reaction cross sections of $x_1$ + A and $x_2$ + A individually, while the other fragments, $x_2$ and $x_1$ respectively, are scattered and not observed. 
\begin{equation}
\frac{E_{{x}_1}}{k_{{x}_1}}\sigma_{R}^{x_1} = \int d\textbf{r}_{{x}_1}d\textbf{r}_{{x}_2} |\hat{S}_{b}(\textbf{r}_{{x}_1}, \textbf{r}_{{x}_2})|^{2}|\chi^{(+)}_{{x}_2}(\textbf{r}_{x_2})|^2
W(\textbf{r}_{{x}_1})|\chi^{(+)}_{{x}_1}(\textbf{r}_{x_1})|^2 , \label{psi4bapp-13}
\end{equation}
\begin{equation}
\frac{E_{{x}_2}}{k_{{x}_2}}\sigma_{R}^{x_2} = \int d\textbf{r}_{{x}_1}d\textbf{r}_{{x}_2} |\hat{S}_{b}(\textbf{r}_{{x}_1}, \textbf{r}_{{x}_2})|^{2}|\chi^{(+)}_{{x}_1}(\textbf{r}_{x_1})|^2
W(\textbf{r}_{{x}_2})|\chi^{(+)}_{{x}_2}(\textbf{r}_{x_2})|^2 . \label{psi4bapp2-13}
\end{equation}

\section{The case of a two-fragment projectile on a two-fragment target}
In the following we treat another four-body breakup problem: the case of two-fragment projectile and two-fragment target. Both projectile and target can break into their two fragments. This is a genuine four-body scattering problem. In principle the formalism of Ref. \cite{CFH2017} can be applied after several modifications. Thus the target is a = b + $x_2$, and the projectile is A = $x_1$ + B. Thus the inclusive spectrum of b will contain breakup of the projectile with $x_1$ interacting with the target a, $x_1$ + a, and the breakup of the target with $x_2$ interacting with the projectile, $x_2$ + A. In principle this process is a complicated four-body reaction. Here, however we take a simpler approach and treat the process as a two three-body problems. As such we have the breakup of the projectile without affecting the target and the breakup of the target without affecting the projectile. In the calculation of the inclusive non-elastic breakup, one would obtain two distinct groups of detected spectator fragments, one related to the target and the other to the projectile. This method would be valuable in the case of a projectile being an exotic, neutron or proton-rich nucleus. 

In the following we consider the reaction $^8$B + d, which leads to  p + (n + $^8$B) $\rightarrow$ p + $^9$B, and p + ($^7$Be + d). We remind the reader that $^8$B is a one proton halo with a halo separation energy of  0.137 MeV. The first reaction results in the neutron capture by a one-proton halo nucleus, while the second reaction results in the incomplete fusion of the core of this halo nucleus with the deuteron target. The inclusive non-elastic proton spectrum can be written as (denoting the proton originating from the radioactive projectile by $p_1$ and that from the deuteron target breakup by $p_2$)
\begin{equation}
\frac{d^{2}\sigma_{p}}{dE_{p}d\Omega_{p}} = \rho(E_{p_2}) \hat{\sigma}_{R}(n + ^8B) + \rho(E_{p_1}) \hat{\sigma}_{R}(d + ^7Be) + \cdot\cdot\cdot
\label{B}
\end{equation}
The first term on the RHS of the above equation contains the neutron capture cross section of the halo nucleus and would be concentrated at higher proton energy (the proton separation energy of the deuteron is 2.22 MeV) in its spectrum, while the second term corresponds to the incomplete fusion, $^7$Be + d, which involves the emission of the halo proton in $^8$B and the collision of its core $^7$Be with the deuteron. This process should dominate the low energy part of the inclusive proton spectrum. 

In the case of a one-neutron halo projectile such as $^{11}$Be or $^{19}$C, with halo neutron separation energies, $E_s$ = 0.501 MeV and $E_s$ = 0.530 MeV, respectively, the same type of reaction will results in an inclusive proton spectrum which should exhibit a now low energy peak related to to the target deuteron breakup at 2.22 MeV, and high energy and weaker peak connected with removing a proton from the tightly bound cores, $^{10}$Be, $^{18}$C.

\begin{equation}
\frac{d^{2}\sigma_{p}}{dE_{p}d\Omega_{p}} = \rho(E_{p_2}) \hat{\sigma}_{R}(n + ^{11}{\rm Be}) + \rho(E_{p_{1}}) \hat{\sigma}_{R}(d +  ^{10}{\rm Be}) + \cdot\cdot\cdot
\label{Be}
\end{equation}
\begin{equation}
\frac{d^{2}\sigma_{p}}{dE_{p}d\Omega_{p}} = \rho(E_{p_2}) \hat{\sigma}_{R}(n + ^{19}{\rm C}) + \rho(E_{p_{1}}) \hat{\sigma}_{R}(d + ^{18}{\rm B}) + \cdot\cdot\cdot
\label{C}
\end{equation}
The cross sections, $\hat{\sigma}_{R}(n + ^{11}$Be), $\hat{\sigma}_{R}(n + ^{19}$C), $\hat{\sigma}_{R}(d + ^{10}$Be), $\hat{\sigma}_{R}(d + ^{18}$B), are given by expressions similar to Eq. (\ref{sigRpar}). One needs the S-matrix elements,  
$\hat{S}_{p_{1}}(\textbf{r}_{p_{1}})$ and $\hat{S}_{p_{2}}(\textbf{r}_{p_{2}})$ in order to evaluate the above cross sections. These matrix elements can be evaluated once appropriate optical potentials for protons on deuteron and on the different halo projectiles are given. Further, optical potentials for the projectile target systems are needed, as well as those for the generation of the participant fragment distorted waves. These are $n + ^{11}$Be, $n + ^{19}$C, $d + ^{10}$Be, $d + ^{18}$B.

For the proton halo nucleus $^8$B, we need similar ingredients: $\hat{S}_{p_{1}}(\textbf{r}_{p_{1}})$ for p + d elastic scattering and  $\hat{S}_{p_{2}}(\textbf{r}_{p_{2}})$ for p + $^8$B. Similarly one needs the d + $^8$B optical potential and the n + $^8$B and 
d + $^7$Be optical potentials. These potentials in principle are known from elastic scattering data.

Once the incomplete fusion cross sections are calculated from fusion theory \cite{CGDH06}, the neutron capture cross sections can be obtained from the general form of the breakup cross sections, Eqs. (\ref{B}, \ref{Be}, \ref{C}).
Thus the Inclusive Non-Elastic Breakup is a potentially powerful method to extract the neutron capture cross section of short-lived radioactive nuclei.

The density of states of the observed proton in Eq.s (\ref{B}), (\ref{Be}) and (\ref{C}) are given by:
\begin{equation}
\rho_{p}(E_p) = \frac{m_p k_p}{(2\pi)^3 \hbar^2}
\end{equation}

In $^8$B + d, due to the low value of the halo proton separation energy, of $E_s$, in the inclusive nonelastic breakup reaction, we expect a low energy peak in the inclusive proton spectrum connected with the incomplete fusion d + $^7$Be, and a higher energy peak connected with the neutron capture n + $^8$B reaction.

In $^{11}$Be + d,  with $E_s$ = 0.5 MeV, we expect a lower energy peak associated with the neutron capture n + $^{11}$Be and a much higher energy peak connected with the incomplete fusion d + $^{10}$Li. The higher energy peak is connected to the proton emitted from the core, $^{10}$Be with a separation energy of $E_s$ = 5 MeV.
Similarly, for $^{19}$C + d: a low energy peak n + $^{19}$C with a higher energy peak d + $^{18}$B.

\section{Outline of the derivation of Eq. (\ref{B}), Eq.(\ref{Be}) and Eq. (\ref{C})}

Here we present an outline of the derivation of Eqs.(\ref{B}, \ref{Be}, \ref{C}). We take the projectile A to be a bound system of two fragments, $x_1$ and B, and the target a as similarly composed of a bound system of two fragments, $x_2$ and b. In this derivation we follow the works of \cite{IAV1985,Austern1987,CFH2017}.

We invoke the spectator model in the sense that the observed fragment is only optically scattered from the projectile or target. Thus we take the Hamiltonian to be
\begin{equation}
H=K_{x_{1}}+K_{x_{2}}+K_{b}+K_{a}+V_{x_{1}x_{2}}+V_{x_{2}A}+V_{x_{2}b}+V_{x_{1}b}+U_{x_{1}A}+U_{bA}.
\end{equation}
The steps to be followed to obtain Eqs. (\ref{B}), Eq.(\ref{Be}) and Eq. (\ref{C}) are lengthy but rest on a  generalization of the case of three-fragment projectile breakup formalison of \cite{CFH2017}, and will be reported elsewhere \cite{BCFH2019}.\\

{\bf Acknowledgements.} 
This work was partly supported by the US-NSF and by the Brazilian agencies, Funda\c c\~ao de Amparo \`a Pesquisa do Estado de S\~ao Paulo (FAPESP), the  Conselho Nacional de Desenvolvimento Cient\'ifico e Tecnol\'ogico  (CNPq) and INCT-FNA project 464898/2014-5. CAB  acknowledges a Visiting Professor support from FAPESP and MSH acknowledges a Senior Visiting Professorship granted by the Coordena\c c\~ao de Aperfei\c coamento de Pessoal de N\'ivel Superior (CAPES), through the CAPES/ITA-PVS program. CAB also acknowledges support by the U.S. NSF Grant No. 1415656 and the U.S. DOE Grant No. DE-FG02-08ER41533.

\end{document}